\documentclass{elsart}
\usepackage{amsmath}
\usepackage{graphicx}
\usepackage{amsfonts}
\usepackage{amssymb}

\input{tcilatex}

\begin{document}

\begin{frontmatter}
\title{The spread of disease with birth and death on networks}

\author{\corauthref{cor1}Jingzhou Liu$^{1}$}
\author{, Yifa Tang$^{2}$}
\author{and Z.R. Yang$^{1}$}

\address{$^{1}$Department of physics, Beijing Normal University,
Beijing,100875, China }%
\address{ $^{2}$Sate Key Laboratory of Scientific and Computing, Chinese Academy of Sciences, Beijing 100080, China}%

 \corauth[cor1]{Corresponding author. E-mail address: liujz606@yahoo.com.cn}%
\begin{abstract}%
\ \ In this paper, we introduce a modified epidemic model on regular
 and scale-free networks respectively. We consider
the birth rate $\delta$, cure rate $\gamma$, infection rate
$\lambda$, $\alpha$ from the infectious disease, and death rate
$\beta$ from other factors. Through mean-field analysis, we find
that on regular network there is an epidemic threshold
$\lambda_{c}$ dependent on the parameters $\delta$,
 $\gamma$, $\alpha$, and $\beta$; while for power law degree distribution
network epidemic threshold is absent in the thermodynamic limit.
The result is the same as that of the standard SIS model. This
reminds us the structure of the networks plays a very important
role in the spreading property of the infectious disease.
\end{abstract}%


\end{frontmatter}

\section{Introduction}

During the past few years, the spread of disease has been one of the focuses
in the field of statistical physics. A a great deal of epidemiological
research work has been done on various networks. Two epidemic models SIS and
SIR have been widely studied[1-9]. In these models, each node of networks
represents an individual and each link is the connection along which the
individuals interact and the disease can spread. For SIS epidemic model,
each individual can exist in two possible states: susceptible (or healthy)
and\ infected. At each time step, each healthy individual can be infected at
rate $\lambda $ if there is one or more infected individuals in its nearest
neighbors. Meanwhile, an infected individual may recover and become
susceptible at rate $\gamma $. The SIR model assumes that individuals can
exist in three possible states: susceptible (or healthy), infected and
removed. The main difference from the SIS model is that once an individual
gets infected, it is removed and can not be infected any more. It is easy to
understand that both the properties of disease and network topology
determine the dynamical behaviors of the disease spreading. Studies of SIS
model and SIR model show that, on regular networks there is an epidemic
threshold $\lambda _{c}$. If the effective spreading rate $\lambda >\lambda
_{c}$, the infection spreads and becomes endemic. Otherwise, the infection
will disappear. While the epidemic threshold is absent on scale-free
networks in the thermodynamic limit[10-12].

From the definitions of SIS and SIR models, we know that both SIS and SIR
models assume the number of individuals to be a constant. However, some
disease may cause enough deaths to influence the population size. So it is
necessary to take the birth and death rates into account. In this paper, we
introduce a modified model, considering the birth rate $\delta $, treatment
rate $\gamma $, infection rate $\lambda $, and two death rates: $\beta $ due
to this infectious disease and $\alpha $ due to other factors, which will be
described later. Our work is to study the influence of above parameters to
the epidemic thresholds on different complex networks.

\section{Model}

We think of our individuals as being spatially distributed on the network $Z$%
. Each site of $Z$\ is empty or occupied by at most one individual. We give
each site a number: 0, 1 or 2. They describe empty state, a healthy
individual occupation and an infected individual occupation respectively.
The state of the system at time $t$ can be described by a set of numbers, 0,
1, 2. That means if the system is in state $A$ and the site $x\in Z$, then $%
A_{t}(x)\in \{0,1,2\}$. Each site can change its state with a certain rate.
An empty site can give birth to a healthy individual at rate $\delta $. A
healthy individual can be infected by contact at rate $\lambda $ if there
are infected individuals in its nearest neighbors, or die at rate $\alpha $
due to other factors. An infected individual can be cured at rate $\gamma $
or die at rate $\beta $ due to this infectious disease. If an individual
dies, there is an empty site left. Of course, each site can also maintain
its state. We define $n_{i}(x,t)$ as the number of the nearest neighbors of
site $x$ in state $i$ at time $t$.%
\begin{align*}
0& \rightarrow 1\text{ at rate }\delta \\
1& \rightarrow 0\text{ at rate }\alpha \text{ } \\
1& \rightarrow 2\text{ at rate }n_{2}\lambda \\
2& \rightarrow 1\text{ at rate }\gamma \\
2& \rightarrow 0\text{ at rate }\beta
\end{align*}

In above expressions, $\delta $, $\alpha $, $\beta $, $\gamma $ and $\lambda 
$ are all non-negative. We assume $\alpha $ is relatively very small. The
expression $n_{2}\lambda $ means that a healthy individual with $n_{2}$
infected nearest neighbors gets infected at rate $n_{2}\lambda $. Not
difficult to see, if $\delta $, $\alpha $ and $\beta $\ equal 0, this model
turns to SIS model; if $\delta $, $\alpha $, and $\gamma $ equal 0, this
model turns to SIR model. If $\alpha $\ and $\gamma $\ equal 0, this model
turns to \textquotedblleft forest fire\textquotedblright , which has been
widely studied\cite{c4}.

\subsection{ Mean-field Method on regular networks}

First, we solve the model by mean-field method on regular network without
the consideration of spatial fluctuation. We use the density $x$\ and $y$\ $%
(x,y\in \lbrack 0,1])$ to replace the numbers of the healthy individuals and
the infected individuals respectively. $"n_{2}\lambda "$ can be replaced as $%
"\lambda \left\langle k\right\rangle y"$, where $\left\langle k\right\rangle 
$ is the average number of the nearest neighbors of one node. The evolution
equations of $x$ and $y$ are governed by:%
\begin{align}
\frac{\partial x}{\partial t}& =(1-x-y)\delta -\alpha x-\lambda \left\langle
k\right\rangle xy+\gamma y  \label{g1} \\
\frac{\partial y}{\partial t}& =\lambda \left\langle k\right\rangle
xy-\gamma y-\beta y  \label{g2}
\end{align}

In Eq.(\ref{g1}), the expression $(1-x-y)$ is the density of empty site. $%
\left\langle k\right\rangle y$ is the probability that the nearest neighbors
of one healthy individual are infectious.

Let $\frac{\partial x}{\partial t}$ $=0$ and $\frac{\partial y}{\partial t}%
=0 $, we get the steady-state solutions:

(I)%
\begin{equation}
x=\frac{\delta}{\alpha+\delta},\text{ }y=0;
\end{equation}

and

(II) 
\begin{equation}
x=\frac{\gamma +\beta }{\lambda \left\langle k\right\rangle },\text{ }y=%
\frac{\delta \lambda \left\langle k\right\rangle -(\delta +\alpha )(\gamma
+\beta )}{\lambda \left\langle k\right\rangle (\delta +\beta )}
\end{equation}

Now, I will do stability analysis. For solution (I), the Jacobean is:%
\begin{equation}
\mathbf{J=}\left( 
\begin{array}{cc}
-\alpha -\delta & \gamma -\delta -\frac{\delta \lambda \left\langle
k\right\rangle }{\delta +\alpha } \\ 
0 & \frac{\delta \lambda \left\langle k\right\rangle }{\delta +\alpha }%
-(\gamma +\beta )%
\end{array}%
\right)
\end{equation}

The determinant and the trace of $\mathbf{J}$: 
\begin{equation}
\left\vert \mathbf{J}\right\vert =-(\alpha +\delta )[\frac{\delta \lambda
\left\langle k\right\rangle }{\delta +\alpha }-(\gamma +\beta )]
\end{equation}

\ 
\begin{equation}
Tr(\mathbf{J})=-\alpha -\delta +\frac{\delta \lambda \left\langle
k\right\rangle }{\delta +\alpha }-(\gamma +\beta )
\end{equation}

Clearly, if $\left\vert \mathbf{J}\right\vert $ \TEXTsymbol{>}0, then $Tr(%
\mathbf{J})<0$, and the solution is stable. So we can get the critical value 
$\lambda_{c}$ of $\lambda$. For simplicity, we let $\delta=1$. Then

\begin{equation}
\lambda _{c}=\frac{(\alpha +1)(\gamma +\beta )}{\left\langle k\right\rangle }
\label{g33}
\end{equation}

If $\lambda<\lambda_{c}$, the solution (I) is stable, and the disease will
die out. Otherwise solution (I) is not stable.

For solution (II), the Jacobean is:

\begin{equation}
\mathbf{J=}\left( 
\begin{array}{cc}
-(\alpha +\delta )-\frac{\delta \lambda \left\langle k\right\rangle -(\delta
+\alpha )(\gamma +\beta )}{\lambda \left\langle k\right\rangle (\delta
+\beta )} & -(\beta +\delta ) \\ 
\frac{\delta \lambda \left\langle k\right\rangle -(\delta +\alpha )(\gamma
+\beta )}{\lambda \left\langle k\right\rangle (\delta +\beta )} & 0%
\end{array}%
\right)
\end{equation}

Considering $y=\frac{\delta \lambda \left\langle k\right\rangle -(\delta
+\alpha )(\gamma +\beta )}{\lambda \left\langle k\right\rangle (\delta
+\beta )}\geq 0$, we also can get $\lambda _{c}$(let $\delta =1$):

\begin{equation}
\lambda _{c}=\frac{(\alpha +1)(\gamma +\beta )}{\left\langle k\right\rangle }
\label{g3'}
\end{equation}

When $\lambda >\lambda _{c}$, the solution (II) is stable, which means that
the disease will pervade the network; otherwise the disease will disappear.
Noticing the expressions(\ref{g33}) and (\ref{g3'}), we find that $\lambda
_{c}$ is a critical parameter. If $\lambda <\lambda _{c}$, the solution (I)
is stable, and the disease will disappear from the network; if $\lambda
>\lambda _{c}$, the solution (II) is stable, and the disease will spread in
the system. From (\ref{g33}) and (\ref{g3'}), it is obvious that $\lambda
_{c}$ is governed by the parameters $\alpha $, $\beta $, $\gamma $, and $%
\left\langle k\right\rangle $. We can increase the treatment rate or
decrease $\left\langle k\right\rangle $ to raise the threshold to prevent
disease from spreading.

\subsection{The spread of disease with treatment on scale-free network}

In the above section, we have studied the epidemic model on regular network.
But the investigations have shown that a large number of systems, such as
internet, world-wide-web, physical, biological, and social network, exhibit
complex topological properties[14-16], particularly scale-free network
feature\cite{albert}. Recent works have examined the spread of computer
viruses on the scale free networks[7,8,10]. The results show that the
intrinsic epidemic threshold is absent in both SIS model and SIR model on
scale-free(SF) networks. In this section, we analyze our modified model on
the scale free networks, of which the degree distribution is $%
p(k)=Cf(k)k^{-\nu }$, where $f(k)$ is the function of $k$. Suppose $x_{k}(t)$
and $y_{k}(t)$ are the density of the healthy and infected nodes with given
degree $k$, and the mean-field equations are[9,10]:%
\begin{align}
\frac{\partial x_{k}(t)}{\partial t}& =\delta (1-x_{k}-y_{k})-\alpha
x_{k}-\lambda kx_{k}\Theta _{k}(y(t))+\gamma y_{k}  \label{g4} \\
\frac{\partial y_{k}(t)}{\partial t}& =\lambda kx_{k}\Theta
_{k}(y(t))-(\gamma +\beta )y_{k}  \label{g5}
\end{align}

where $\Theta _{k}(y(t))$ stands for the probability that an edge emanating
from a node of degree $k$ points to an infected site, $\Theta
_{k}(y(t))=\sum_{k^{^{\prime }}}p(k^{^{\prime }}/k)y_{k^{^{\prime }}}(t)$,
where $p(k^{^{\prime }}/k)$ is the probability that a node with $k$ degree
points to a node with $k^{^{\prime }}$ degree. For uncorrelated networks\cite%
{pk1}, $p(k^{^{\prime }}/k)=k^{^{\prime }}p(k^{^{\prime }})/\left\langle
k\right\rangle $, which means that the probability that a node points to a
node with $k^{^{\prime }}$ degree is proportional to its degree and the
degree distribution $p(k^{^{\prime }})$, and $\left\langle k\right\rangle $\
is the normalization factor. From the definition of $\Theta _{k}(y(t))$, we
find that it is independent of $k$ for uncorrelated networks\cite{pk1}:%
\begin{equation}
\Theta _{k}(y(t))=\Theta (y(t))=\left\langle k\right\rangle
^{-1}\sum_{k^{\prime }}k^{\prime }p(k^{\prime })y_{k^{\prime }}(t)
\label{g6}
\end{equation}

Let $\frac{\partial x_{k}(t)}{\partial t}=0$ and $\frac{\partial y_{k}(t)}{%
\partial t}=0$, we can get stationary solution:%
\begin{align}
x_{k}& =\frac{\gamma +\beta }{(1+\alpha )(\gamma +\beta )+(1+\beta )\lambda
k\Theta }  \label{g7} \\
y_{k}& =\frac{\lambda k\Theta }{(1+\alpha )(\gamma +\beta )+(1+\beta
)\lambda k\Theta }  \label{g8}
\end{align}

In the above expression, we have let $\delta=1.$.

Substituting the expression (\ref{g8}) into (\ref{g6}), we get
self-consistent equation of $\Theta $:%
\begin{equation}
\Theta =\frac{1}{(1+\beta )\left\langle k\right\rangle }\sum_{k}p(k)\frac{%
\lambda ^{\prime }k^{2}\Theta }{1+\lambda ^{\prime }k\Theta }=\frac{1}{%
(1+\beta )\left\langle k\right\rangle }\left\langle \frac{\lambda ^{\prime
}k^{2}\Theta }{1+\lambda ^{\prime }k\Theta }\right\rangle  \label{g9}
\end{equation}

where $\lambda^{\prime}=\frac{1+\beta}{(1+\alpha)(\gamma+\beta)}\lambda$.

We can see that $\Theta =0$ is a solution of Eq.(\ref{g9}). To allow a
nonzero solution $\Theta (\Theta \in (0,1])$ of Eq.(\ref{g9}), the following
inequality must be assumed:%
\begin{equation}
\left( \frac{1}{(1+\beta )\left\langle k\right\rangle }\left\langle \lambda
^{\prime }k^{2}\Theta \right\rangle \right) \geq 1  \label{g10}
\end{equation}

From Eq. (\ref{g10}), we get the threshold value of $\lambda ^{\prime }$:

\begin{equation}
\lambda _{c}^{\prime }=(1+\beta )\frac{\left\langle k\right\rangle }{%
\left\langle k^{2}\right\rangle }  \label{g11}
\end{equation}

\bigskip where $\left\langle k^{2}\right\rangle =\sum_{k}k^{2}p(k)$, then:%
\begin{equation}
\lambda _{c}=(1+\beta )(\gamma +\alpha )\frac{\left\langle k\right\rangle }{%
\left\langle k^{2}\right\rangle }  \label{66}
\end{equation}

From (\ref{66}), we can see that $\lambda _{c}$ is dependent on $\gamma $, $%
\alpha $, $\beta $, and $\frac{\left\langle k\right\rangle }{\left\langle
k^{2}\right\rangle }$. If $\lambda >\lambda _{c}$, the disease will spread
on the networks, otherwise the disease will die out. We now discuss $\lambda
_{c}$ for different $f(k)$.

(I) For $f(k)=\delta _{k,k_{c}}$, then $p(k)=Ck^{-\upsilon }\delta
_{k,k_{c}}(k_{c}\geq 2)$. The network is homogeneous, $\left\langle
k\right\rangle =k_{c}$, $\left\langle k^{2}\right\rangle =k_{c}^{2}$, so%
\begin{equation}
\lambda _{c}=\frac{(1+\alpha )(\gamma +\beta )}{k_{c}}  \label{100}
\end{equation}

Clearly, there is a nonzero threshold $\lambda _{c}$, in agreement with the
result on the regular network(see Eq. (\ref{g3'})). When $\lambda >\lambda
_{c}$, there is a nonzero $\Theta =\frac{\lambda \left\langle k\right\rangle
-(\delta +\alpha )(\gamma +\beta )}{\lambda \left\langle k\right\rangle
(1+\beta )}$of Eq.(\ref{g9}). $\lambda _{c}$ is an increasing function of $%
\gamma $, $\alpha $ and $\beta ,.$ We can increase the threshold $\lambda
_{c}$ by increasing the rate of treatment $\gamma $ and decreasing $k_{c}$.

(II) For $f(k)=1$, the network is scale free with a power law degree
distribution $p(k)=Ck^{-\upsilon }(\upsilon \in (2,3])$, then 
\begin{align}
\left\langle k\right\rangle & =\sum_{k=m}^{+\infty }kp(k)\simeq C\frac{1}{%
\upsilon }m^{2-\upsilon } \\
\left\langle k^{2}\right\rangle & =\sum_{k=m}^{+\infty }k^{2}p(k)\simeq
\int_{m}^{\infty }k^{2-\upsilon }dk
\end{align}

As $(2-\upsilon )$ is bigger than $-1$, so $\left\langle k^{2}\right\rangle $
is divergent, and $\frac{\left\langle k\right\rangle }{\left\langle
k^{2}\right\rangle }\rightarrow 0$, then $\lambda _{c}\rightarrow 0$ for $%
k\rightarrow \infty $. Therefor the threshold is absent. Thus the treatment
is of no effect to the disease.

(III) For $f(k)=e^{-k/k_{c}}$, $f(k)$ decreases rapidly for $k>k_{c}$, the
network is a finite size scale free network\cite{c2}. Then%
\begin{align}
\lambda _{c}& =(1+\alpha )(\gamma +\beta )\frac{\left\langle k\right\rangle 
}{\left\langle k^{2}\right\rangle }  \notag \\
& =(1+\alpha )(\gamma +\beta )\frac{\sum_{k}k^{1-\upsilon }e^{-k/k_{c}}}{%
\sum_{k}k^{2-\upsilon }e^{-k/k_{c}}}  \notag \\
& =(1+\alpha )(\gamma +\beta )k_{c}^{-1}\frac{\Gamma (-\upsilon ,m/k_{c})}{%
\Gamma (1-\upsilon ,m/k_{c})}  \label{g}
\end{align}

Where $\Gamma(x,y)$ is the incomplete gamma function. The threshold $%
\lambda_{c}$ is nonzero for finite $k_{c}$. Without surprise, the threshold
is an increasing function of $\gamma$, $\alpha$, $\beta$. Comparing (\ref{g}%
) with (\ref{100}), one can see, for network with degree distributions $%
p(k)=Ck^{-\upsilon}e^{-k/k_{c}}$ $\lambda_{c}$ is also very small. From
(I)-(III), we find that the network degree distribution, in some sense,
determines the spread of infectious disease.

\section{Conclusion}

To summary, we have suggested an epidemic model with birth rate and death
rate. Through mean-field analysis, we find that on regular network the
epidemic threshold is an increasing function of treatment rate $\gamma $,
death rates $\alpha $ and $\beta $; while for power law degree distribution
network epidemic threshold is absent in the thermodynamic limit, so that the
treatment thus is of no effect to the disease, which is the same as the
result of standard SIS model. So to prevent the infectious disease spreading
in the "networks", apart from increasing the cure rate, we should pay more
attention to the structure of "networks". We should point out that we do not
analyze specific disease in our model. For a specific disease, there is a
great need to analyze the disease through modelling and comparing the
epidemic model with real data.

\begin{center}
ACKNOWLEDGMENT
\end{center}

This work was supported by the National Science Foundation of China under
Grant No. 10175008. We acknowledge the support from The State Key Laboratory
of Scientific and Engineering Computing (LSEC), Chinese Academic of Sciences.

\end{document}